\documentclass[reprint, amsmath,amssymb, aps, superscriptaddress]{revtex4-1}

\usepackage{graphicx}
\usepackage{dcolumn}
\usepackage{bm}
\usepackage{color}

\begin{document}

\preprint{UTCCS-P-70}

\title{Conformal Theories with an IR cutoff}


\author{K.-I. Ishikawa}
\affiliation{Graduate School of Science, Hiroshima University,Higashi-Hiroshima, Hiroshima 739-8526, Japan}

\author{Y. Iwasaki}
\affiliation{Center for Computational Sciences, University of Tsukuba,Tsukuba, Ibaraki 305-8577, Japan}

\author{Yu Nakayama}
\affiliation{California Institute of Technology,  Pasadena, CA 91125, USA}

\author{T. Yoshie}
\affiliation{Center for Computational Sciences, University of Tsukuba,Tsukuba, Ibaraki 305-8577, Japan}

\date{\today}

\begin{abstract}
We give a new perspective on the dynamics of conformal theories realized in the  SU($N$) gauge theory,
when the number of flavors $N_f$ is within the conformal window.
Motivated by the RG argument on conformal theories with a finite IR cutoff $\Lambda_{\mathrm{IR}}$,
we conjecture that
the propagator of a meson $G_H(t)$ on a lattice behaves  at large $t$ as a power-law corrected Yukawa-type decaying form $G_H(t) = \tilde{c}_H\, \exp{(-\tilde{m}_H t)}/t^{\alpha_H}$  instead of the exponentially decaying form $c_H\exp{(-m_H t)}$, in the small quark mass region where  $m_H \le c\, \Lambda_{\mathrm{IR}}$: $m_H$ is the mass of the ground state hadron in the channel $H$ and $c$ is a constant of order 1. 
The transition between the ``conformal region'' and the ``confining region" is a first order transition.
Our numerical results verify the predictions for the $N_f=7$ case and the $N_f=16$ case in the SU($3$) gauge theory with the fundamental representation.
\end{abstract}

\pacs{12.38.Gc, 11.25.Hf}
\maketitle


Conformal field theories are ubiquitous in nature and play important roles not only in particle physics beyond the standard model but also in condensed matter physics. Nonperturbative understanding of their dynamics in four-dimensional space-time is ardently desired. In this article, we give a new perspective on the dynamics of conformal theories realized in the  SU($N$) gauge theory
when the number of flavors $N_f$ is within the conformal window  \cite{Banks1982} by studying meson propagators with a finite infrared (IR) cutoff. Some preliminary results have been presented in \cite{iwa2012}.

Our general argument that follows can be applied to any gauge theories with arbitrary representations as long as they are in the conformal window, but to be specific, we focus on SU(3) gauge theories with $N_f$ fundamental fermions (``quarks"). We define the conformal field theory in a constructive way \cite{review}: We employ the Wilson quark action and the standard one-plaquette gauge action on the Euclidean lattice of the size $N_x=N_y=N_z=N$ and $N_t=r N$ with  aspect ratio $r$. We impose periodic boundary conditions except for  an anti-periodic boundary condition in the
time direction for fermion fields. We eventually take the continuum limit by sending the lattice space $a \rightarrow 0$ with $N \rightarrow \infty$ 
keeping $L =N \, a$  fixed. 
When  $L=$ finite, the continuum limit defines a theory with an IR cutoff.
The theory is defined by two parameters; the bare coupling constant $g_0$ and the bare degenerate quark mass $m_0$ at ultraviolet (UV) cutoff.
We also use, instead of $g_0$ and $m_0$, 
$\beta={6}/{g_0^2}$
and 
$K= 1/2(m_0a+4)$. For  a later purpose, we define the quark mass $m_q$
through Ward-Takahashi identities
with renormalization constants being suppressed \cite{iwa2012}.

Let us quickly remind ourselves of the renormalization group (RG) flow when $N_f$ is in the conformal window. One important fact is finite size lattices in computer simulation always introduce an IR cutoff $\Lambda_{\mathrm{IR}} \sim 1/(N a)$.
If the IR cutoff were zero,
when quarks have tiny masses,  the RG trajectory would stay close to the critical line, approaching the IR fixed point and finally would pass away from the IR fixed point to infinity. Therefore the IR behavior is governed by the ``confining region''. Only on the massless quark line the scale invariance is realized at the IR fixed point. See the left panel of FIG.~1.

When the cutoff $\Lambda_{\mathrm{IR}}$ is finite, the RG flow from UV to IR does stop evolving at the scale $\Lambda_{\mathrm{IR}}$.
When the typical mass scale (e.g. that of a meson) $m_H$ is smaller than $\Lambda_{IR}$, it is in the ``conformal region''.
On the other hand, when $m_H$ is larger than $\Lambda_{\mathrm{IR}}$, the flow passes  away from the IR fixed point to infinity  with relevant variables integrated out, thus being in the  ``confining region''. See the right panel of FIG.~1.

This scenario implies that
when physical quantities at IR (e.g. hadron masses) are mapped into a diagram in terms of physical parameters at UV (e.g. the bare coupling constant and the bare quark mass), there will be gaps in the physical quantities along the boundary between the two phases.
There the phase transition will be a first order transition.

To make the above RG argument more concrete, we will study the propagator of the local meson operator 
\begin{equation}
G_H(t) = \sum_{x} \langle \bar{\psi}\gamma_H \psi(x,t) \bar{\psi} \gamma_H \psi(0) \rangle \ .
\end{equation}
When the theory is in the ``confining region", it decays exponentially at large $t$ as
\begin{equation} G_H(t) = c_H \, \exp(-m_H t)\label{exp}\end{equation}
due to the physical one-particle pole, where $m_H$ is the mass of the ground state hadron in the channel $H$.

\begin{figure*}[htb]
\includegraphics[width=7.5cm]{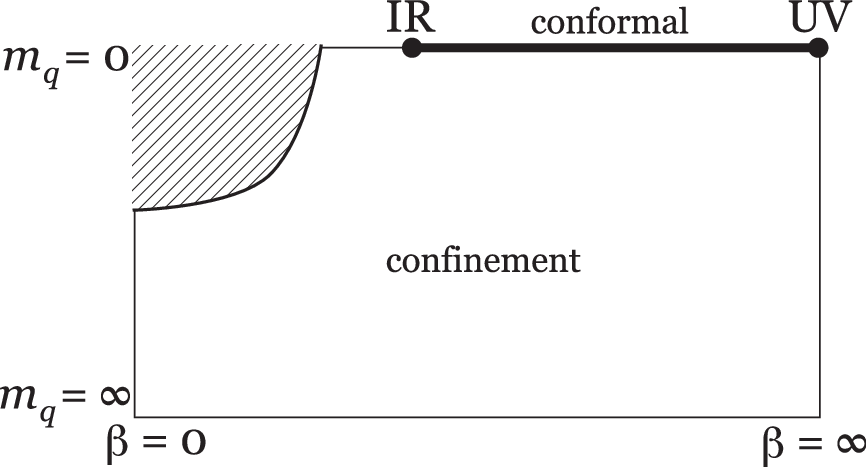}
\hspace{1.5cm}
\includegraphics[width=7.5cm]{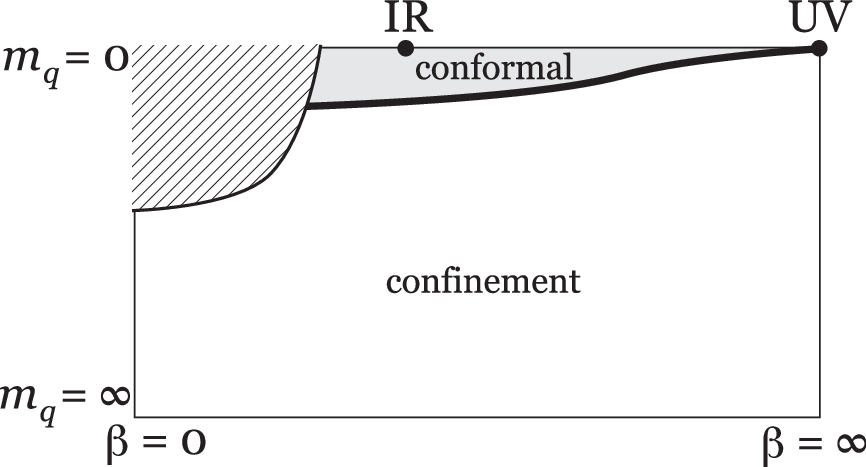}
\caption{
The phase diagram predicted from the RG argument: (left: for $\Lambda_{\mathrm{IR}}=0$) and (right: for $\Lambda_{\mathrm{IR}}=$ finite).
The shaded strong coupling region for small quark masses does not exist in the $\beta - m_q$ plane {\cite{iwa2012}}.
}
\label{PDIR}
\end{figure*}

In contrast, we claim that in the ``conformal region" 
 defined by
\begin{equation} 
m_H   \leq c \,  \Lambda_{\mathrm{IR}},
\label{critical mass}
\end{equation}
where $c$ is a constant of order 1 which we will determine, the propagator $G(t)$ behaves at large $t$  as
 \begin{equation}
G_H(t) = \tilde{c}_H\\ \frac {\exp(-\tilde{m}_Ht)}{t^{\alpha_H}},
\label{yukawa type}
 \end{equation}
which is a power-law corrected Yukawa-type decaying form instead of the exponential decaying form (eq.(\ref{exp})) observed in the ``confining region".
 We further claim that the boundary between the ``conformal region'' and the ``confining region'' is a first order transition.

We note that the behavior eq.(\ref{yukawa type}) is proposed based on the AdS/CFT correspondence with a softwall cutoff in the literature \cite{Cacciapaglia:2008ns}.
 The meson propagator in the momentum space has a cut instead of a pole: 
$G_H(p)= 1/(p^2+\tilde{m}_H^2)^{1-\alpha_H}$.
The propagator in the position space (after space integration)
takes the form eq.(\ref{yukawa type}) in the limit  $t \, \tilde{m}_H \gg 1$.

We distinguish $\tilde{m}_H$ in eq.(\ref{yukawa type}) from the pole mass $m_H$ in eq.(\ref{exp}). Eq.(\ref{critical mass}) thus
means the lower limit of $m_H$ is $c\, \Lambda_{\mathrm{IR}}.$

In the continuum limit  with $L= \infty$ (i.e. $\Lambda_{\mathrm{IR}} = 0$), the propagator on the massless quark line takes the form
\begin{equation} G_H(t) = \tilde{c} \, \frac {1}{t^{\alpha_H}},\end{equation}
consistent with $\tilde{m}_H=0$ limit of eq. (\ref{yukawa type}).
If we take the coupling constant $g_0=g^{*}$ at the UV cutoff, $\alpha_H$ takes a constant value, and the RG equation demands 
\begin{equation} \alpha_H=3 - 2 \gamma^{*},\label{anoma}\end{equation}
for the pseudo-scalar (PS) channel
with $\gamma^{*}$ being the anomalous mass dimension $\gamma$ at $g=g^{*}$.
The theory is scale invariant (and shown to be conformal
invariant within perturbation theory \cite{Polchinski:1987dy}.
See also e.g. \cite{Nakayama:2010zz} and references therein from AdS/CFT approach).
When  $0 \le g_0 <  g^*$, $\alpha_H$ depends slowly on $t$ as a solution of the  RG equation. In the IR limit $t\to \infty$, we must retain $\alpha_H(t) \to 3-2\gamma^{*}$.


The continuum limit with $L=$ finite defines a continuum theory on $\mathbf{T}^3\times \mathbf{S}^1$ which corresponds to a compact three-torus at finite temperature.
The IR cutoff $\Lambda_{\mathrm{IR}} \sim 1/L$ is finite. The propagator $G_H(t)$ behaves at large $t$ as a power-law corrected Yukawa type decaying form 
eq.(\ref{yukawa type}). 
The exponent $\alpha_H$ in $t\to \infty$ with $t \,\tilde{m}_H\ \ll 1$ takes the universal formula $3-2\gamma^{*}$ while with $ t \,\tilde{m}_H \gg 1$ it takes a value depending on the dynamics,
from which we can estimate the mass anomalous dimensions 
 (see  \cite{iwa2012}).

 \begin{figure*}[htb]%
 \begin{center}
\includegraphics[width=7.8cm]{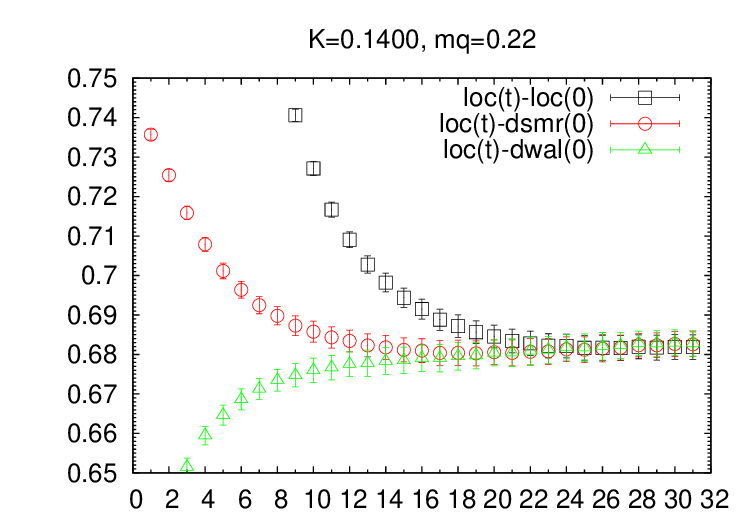}
\hspace{1.5cm}
\includegraphics[width=7.8cm]{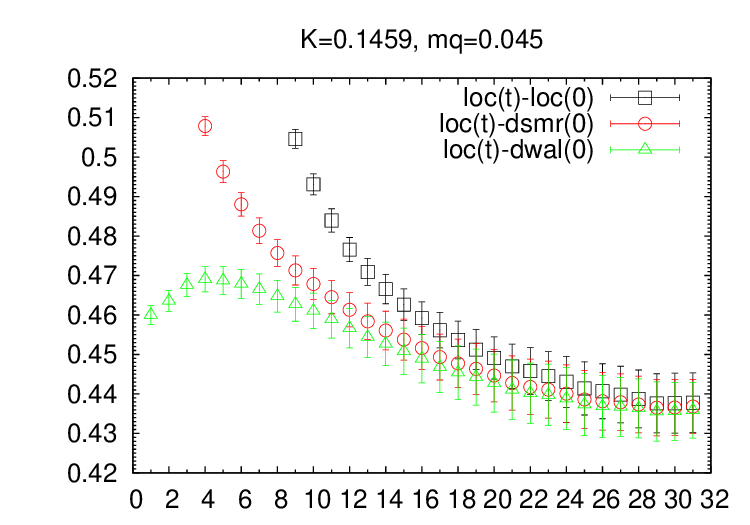}
  \label{exp_vs_yukawa}
  \caption{The effective mass for $N_f=7$ (left: for $K=0.1400$; exponential decay) 
   and (right: for $K=0.1459$; power-law corrected Yukawa decay):
With three types of sources;  the local-sink local-source (black squares), 
local-sink doubly-smeared-source (red circles) and 
local-sink doubly-wall-source (green triangles).
   }
  \end{center}
\end{figure*}

\begin{figure*}[bht]%
  \includegraphics[width=7.8cm]{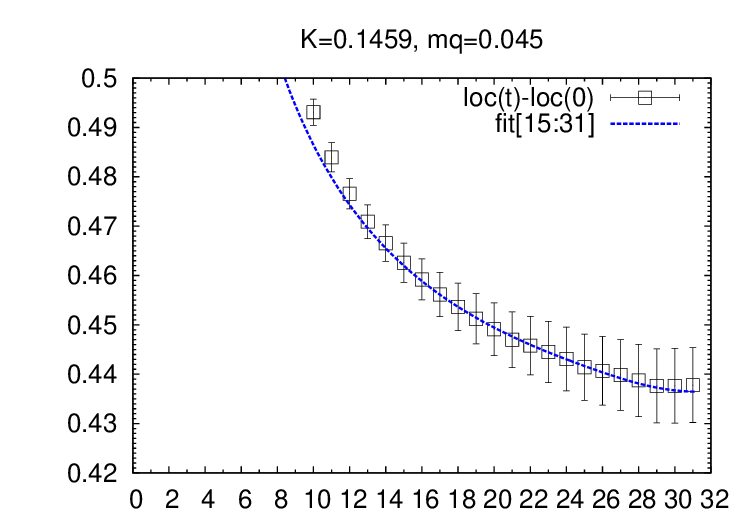}
  \hspace{1.5cm}
    \includegraphics[width=7.8cm]{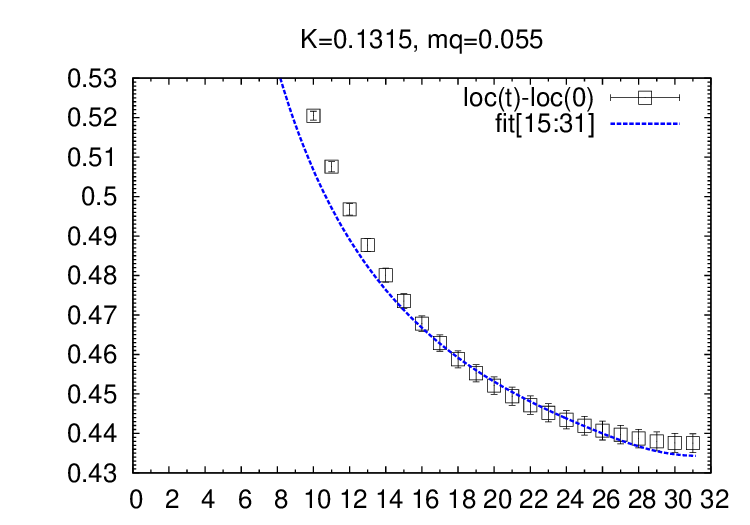}
        \label{expnf7nf16}
        \caption{ 
        The effective mass plots for local-sink local-source case and fits by power-law corrected Yukawa type decay (left: for $N_f=7$ and $K=0.1459$) 
   and (right: for $N_f=16$ and $K=0.1315$).}
\end{figure*}

Now let us discuss the results of our numerical calculations.
We have performed simulations in the  $N_f=7$ and $N_f=16$ cases,
which we conjecture \cite{iwa2004}, are the boundaries of the conformal window. 
The algorithms we employ are the blocked HMC algorithm \cite{Hayakawa:2010gm} for $N_f=2\, N$ and the RHMC algorithm \cite{Clark:2006fx} for $N_f=1$ in the case $N_f=2\, N +1$.

We specify the coupling constant $\beta=6.0$ for $N_f = 7$ and $\beta=11.5$ for $N_f=16,$
taking account of the fact that the IR fixed point for $N_f=16$ is $\beta=11.48$ in two-loop approximation. 
We use the lattices of size
$16^3\times 64$ and $24\times 96,$
and vary the hopping parameter $K$ so that the quark mass takes the value from 0.40 to 0.0: We simulate with
15 hopping parameters on the $16^3\times 64$ lattices, and 5 hopping parameters on  the $24^3\times 96 $ lattices. 

We choose the run-parameters in such a way that the acceptance of the global metropolis test is about $70\%.$
The statistics are 1,000 MD trajectories for thermalization and 1,000 MD trajectories or 500 MD trajectories for the measurement.
We estimate the errors by the jack-knife method
with a bin size corresponding to 100 HMC trajectories.

Let  us first discuss the results for the $N_f=7$ case on the $16^3\times 64$ lattices.  We define the effective mass $m_H(t)$  by $\frac{\cosh(m_H(t)(t-N_t/2))}{\cosh(m_H(t)(t+1-N_t/2))}=\frac{G_H(t)}{G_H(t+1)}$.
FIG.~2 \ref{exp_vs_yukawa} shows 
the $t$ dependence of the effective mass
for the PS channel with three types of sources.
On the left panel, we see the clear plateau of the effective mass at $t=22\sim 31$ when quark mass is relatively large; $m_q=0.25$ ($K=0.1400$). 
On the other hand, on the right panel, we see the effective mass is slowly decreasing without no plateau up to $t=31$ when the quark mass is small; $m_q=0.045$ ($K=0.1459$), suggesting the power-law correction.

The effective masses for all cases with $m_q \le 0.172$ ($K \ge 0.1415$) exhibit a similar behavior to the $K=0.1459$ case.
We show the power-law corrected fit for the local-local data in the $K=0.1459$ case with the fitting range $t=[15:31]$ on the 
left panel in FIG.~3.
The fit with $\alpha_H=0.8(1)$ reproduces the date very well.

In the case of $N_f=16$, the effective masses on the $16^3\times 64$ lattices 
exhibit the power-law corrected Yukawa-type decaying form 
for all cases when $m_q \le 0.0978$ ($K \ge 0.130$).
We show the  power-law corrected fit for the local-local data in the $K=0.1315$
 case with the fitting range $t=[15:31]$
 on the  right panel in FIG.~3. 
The fit with $\alpha_H=1.15(2)$ reproduces the data well.

We also report that the data on $24^3\times 96$ lattices for both cases of $N_f=7$ with $K \ge 0.1459$ and $N_f=16$ with
$K \ge 0.130$ clearly exhibit the power-law corrected Yukawa type decaying form.
We obtain $\alpha_H = 0.54(17)$  for $K=0.1459$ in $N_f = 7$ and $\alpha_H =1.57(28)$
 for $K=0.130$  in $N_f = 16$, respectively.
Thus we have confirmed that the propagators behaves as the power-law corrected Yukawa type decaying form, when the quark mass is small enough.
We note in all cases the rho meson mass is degenerate with the pion mass within the one standard deviation. This implies the phase is a chiral symmetric phase.

In the remaining part of the article, we try to identify the boundary between the ``conformal region'' and the ``confining region''
more precisely by investigating the transition region carefully
with a small step of the value of $K$
  for both of the $N_f=7$ and $N_f=16$ cases.

We present the results for $m_q$ and $m_{PS}$ (or $\tilde{m}_{PS}$) in the $N_f=7$ case in FIG.~4.
 We first note that the quark mass $m_q$ denoted by black points and line on the left panel is excellently proportional to $1/K$ in the whole region from $0.006$ to $0.555$. 
 
 \begin{figure*}[hbt]
 \begin{center}
\includegraphics[width=17.0cm]{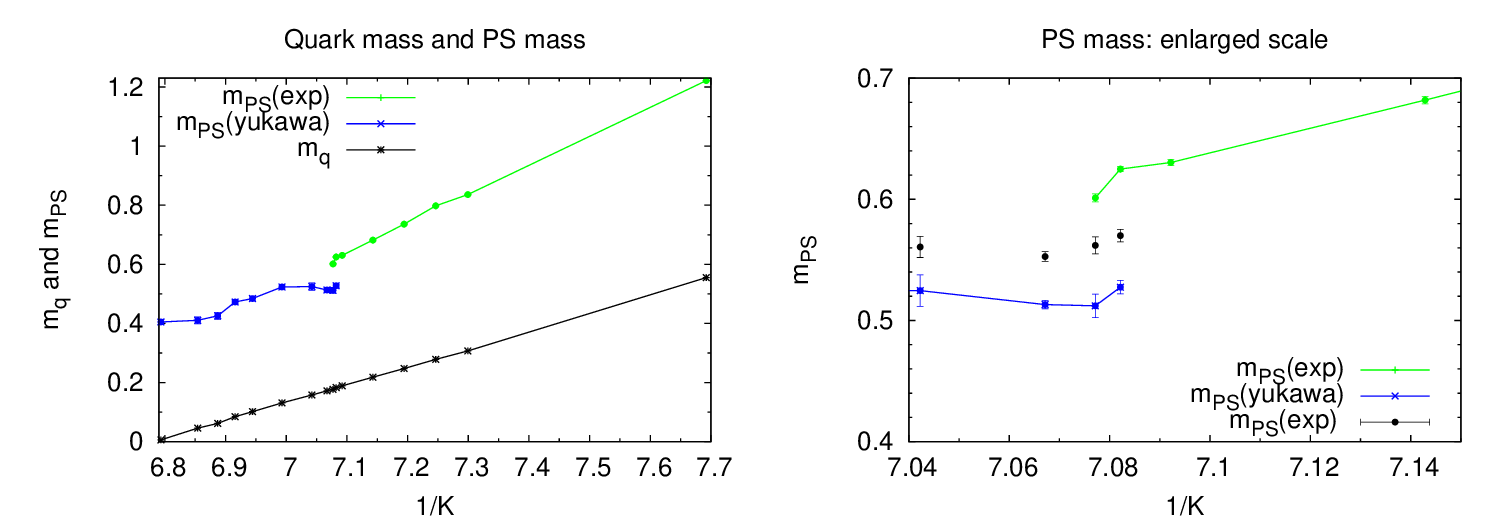}
\caption{ 
$m_q$ and $m_{PS}$ (or $\tilde{m}_{PS}$) vs. $1/K$ for the range $0.130 \le K \le 0.1472$.
The transition region is enlarged on the right panel. See the text for the symbols.
}
\label{mass}
\end{center}
\end{figure*}

For the propagators of the PS meson,
we observe the clear transition from the exponentially decaying form to the power-law corrected Yukawa-type decaying form
at $K =0.1412 \sim 0.1413$.
For $K \le 0.141$, we present $m_{PS}$ obtained from the fit to the exponentially decaying form with the fitting region [28:31] by green
points and line.
For $K \ge 0.1415$, we present $\tilde{m}_{PS}$ obtained 
from the fit to the power-law corrected Yukawa type decaying form with fitting range
[15:31] by blue points and line.

The transition region is enlarged on the right panel in FIG.~4.
The values of $m_{PS}$ and  $\tilde{m}_{PS}$ are different  in the limit $K = 0.1412$ from smaller $K$ and from larger $K$.
It should be noted that even if we had ignored the power-law corrections in Yukawa type decay and had estimated the $m_{PS}$ (plotted in FIG.~4 by black points) by assuming exponential decay
with the fitting range [28:31], although no plateau is seen as shown on the right panel of Fig  2 and in Fig. 3,
there would have been a gap at the transition point.

It is worthwhile to stress that
at $K=0.1412$ and $0.1413$ we observe two states, 
depending on the initial sate.
The existence of two states persists at least with 500 MD trajectories.
We also note that the exponent $\alpha_H$ does not vanish toward the transition point; it rather increases up to $\alpha_H = 1.1 \sim 1.3.$ 
Thus the transition is a first order transition at $K=0.1412 \sim 0.1413.$
These facts imply that there indeed exists a ``conformal region'' which is separated by a gap of physical quantities from the ``confining region" at finite quark mass. 

The $m_q$ dependence of $\tilde{m}_{PS}$ is rather complicated. Once $\tilde{m}_{PS}$ becomes small at the transition point, it slightly increases as $m_q$ decreases, and then decreases at $K \le 0.143.$
Apparently the small $\tilde{m}_{PS}$ region suffers from finite size effects.
To verify the scaling relation \cite{miransky} for $\tilde{m}_{PS}$ in terms of $m_q$ we have to control finite size effects.
 
Taking $m_{PS} = 0.601 \sim 0.625$ with  $m_q =0.177\sim 0.183$ ($K=0.1413 \sim 0.1412$),
as the critical mass
in eq.(3), we estimate $c=2.16 \sim 2.25,$
with our working definition of $\Lambda_{\mathrm{IR}}=2\, \pi (N^3 \times N_t)^{-1/4}$.

In the $N_f=16$ case we have observed similar results. 
The $m_q$ dependence of $\tilde{m}_{PS}$ is similarly complicated.
The transition occurs at $m_{PS}= 0.513 \sim 0.539$ with $m_q=0.237 \sim 0.244$ ($K=0.1255 \sim 0.125$)
from which we estimate $c=1.85 \sim1.94.$

In the cases of $24^3\times 96$ lattices for both $N_f=7$ and $N_f=16$ cases, we have also observed the transition
 : In $N_f=7$, we find exponential decay for $m_q \ge 0.061$ ($K \le 0.1452$)  and
power-law corrected Yukawa type decay for  $m_q \le 0.045 $ ($K \ge 0.1459$). In $N_f=16$, we find power-law corrected Yukawa type decay for $m_q \le 0.0978$ ($K \ge 0.130$).
The critical mass decreases compared with the $16^3\times 64$ lattices as expected since the IR cutoff decreases.

We conclude the paper with two main goals achieved in our study. We have given evidence that  (1): The meson propagator shows a transition from an exponentially decaying form
to a power-law corrected Yukawa-type decaying form 
at the critical hadron mass whose value is specified by the IR cutoff. (2): The transition is first order. 
The existence of the conformal region we proposed  gives a necessary (but not sufficient) condition for the
conformal window, and
the result for $N_f=7$ is consistent with our conjecture that $N_f=7$ is within the conformal window.
However we understand that we need more evidence to conclude that.

In an accompanying publication \cite{else}, we will discuss the physical interpretation of the power-law corrections in terms of the dynamics of the underlying conformal field theory by taking the continuum limit. We also hope to report the physical applications of our findings in the near future.

\section*{Acknowledgments}
We would like to express our gratitude to T. Yanagida for making a chance to start this collaboration.
We thank K. Kanaya for his help in preparing the manuscript.
The calculations were performed with HA-PACS computer at CCS, University of Tsukuba and SR16000
at KEK. We would like to thank members of CCS and KEK for their strong support for this work.

\end{document}